\documentclass[10pt,conference]{IEEEtran}

\IEEEoverridecommandlockouts
\usepackage{caption}
\usepackage{subfigure}
\usepackage{graphics} 
\usepackage{epsfig} 
\usepackage{amsfonts,amssymb}
\usepackage{multirow}
\usepackage{amsmath}
\usepackage{url}
\graphicspath{{figure/}}
\usepackage{subfigure}
\usepackage{graphics}
\usepackage{multirow}
\usepackage{xcolor}
\usepackage{color}

\usepackage{algorithm}
\usepackage{algpseudocode}
\usepackage{cite}

\begin{document}

\title{\huge Leveraging Bi-Directional Channel Reciprocity for Robust Ultra-Low-Rate Implicit CSI Feedback with Deep Learning }

\author{
		\IEEEauthorblockN{
            Zhenyu~Liu\IEEEauthorrefmark{1},
            Yi~Ma\IEEEauthorrefmark{1},
            Rahim~Tafazolli\IEEEauthorrefmark{1},
            Zhi~Ding\IEEEauthorrefmark{2}}
            
		\IEEEauthorblockA{\IEEEauthorrefmark{1}6GIC, Institute for Communication Systems, University of Surrey, Guildford, UK, GU2 7XH}
		\IEEEauthorblockA{\IEEEauthorrefmark{2}Department of Electrical and Computer Engineering, University of California, Davis, California, USA}
        { Emails: (zhenyu.liu, y.ma, r.tafazolli)@surrey.ac.uk, zding@ucdavis.edu }}

\maketitle

\begin{abstract}
Deep learning-based implicit channel state information (CSI) feedback has been introduced to enhance spectral efficiency in massive MIMO systems. Existing methods often show performance degradation in ultra-low-rate scenarios and inadaptability across diverse environments. In this paper, we propose Dual-ImRUNet, an efficient uplink-assisted deep implicit CSI feedback framework incorporating two novel plug-in preprocessing modules to achieve ultra-low feedback rates while maintaining high environmental robustness. First, a novel bi-directional correlation enhancement module is proposed to strengthen the correlation between uplink and downlink CSI eigenvector matrices. This module projects highly correlated uplink and downlink channel matrices into their respective eigenspaces, effectively reducing redundancy for ultra-low-rate feedback. Second, an innovative input format alignment module is designed to maintain consistent data distributions at both encoder and decoder sides without extra transmission overhead, thereby enhancing robustness against environmental variations. Finally, we develop an efficient transformer-based implicit CSI feedback network to exploit angular-delay domain sparsity and bi-directional correlation for ultra-low-rate CSI compression. 
Simulation results demonstrate successful reduction of the feedback overhead by 85\% compared with the state-of-the-art method and robustness against unseen environments.

\end{abstract}
\IEEEpeerreviewmaketitle

\section{Introduction}
With the establishment of a stable system architecture for fifth-generation (5G) networks, wireless standards are now exploring beyond 5G (B5G) systems toward sixth-generation (6G) technologies~\cite{3gpp2022,3gpp2023csi}. Among the enabling technologies, advanced massive multiple-input multiple-output (MIMO) systems play a pivotal role due to their potential to significantly improve both spectral and energy efficiency. However, deploying large antenna arrays and numerous subcarriers in massive MIMO results in substantial overhead for channel state information (CSI) feedback, creating a notable challenge in frequency division duplex (FDD) systems. Consequently, the design of efficient downlink CSI feedback mechanisms has become critically important.

Deep learning (DL)-based CSI feedback approaches~\cite{cui2022, full_wang2023, bifull_liu2019, bifull_2024xu} have demonstrated exceptional performance in terms of compression efficiency and reconstruction accuracy. By effectively exploiting spatial and spectral correlations~\cite{cui2022, full_wang2023}, as well as bi-directional channel correlations~\cite{bifull_liu2019, bifull_2024xu}, these methods significantly outperform conventional CSI feedback schemes. Furthermore, the recognition of DL-based CSI feedback as a key use case in the 3GPP TR 38.843~\cite{3gpp2023csi} highlights its practical potential. Nevertheless, existing DL-based methods predominantly address the full CSI feedback scenario, where user equipment (UE) transmits the entire downlink channel matrix to the base station (BS). In contrast, practical deployments of 5G New Radio (NR)~\cite{3gpp2022} primarily adopt implicit CSI feedback, where UEs only report channel eigenvectors. Given that commercial systems are already based on implicit feedback frameworks, replacing them entirely is impractical. Hence, developing effective DL-based solutions tailored specifically for eigenvector compression and feedback remains crucial.

Recent studies on DL-based implicit CSI feedback~\cite{eigennet_liu, implicit_2022chen, implicit_2024gao, mixeddata2024_2} have successfully aligned  with 3GPP-defined codebooks and exhibited superior performance compared to enhanced Type II codebooks~\cite{3gpp2022}. For instance, EVCsiNet~\cite{eigennet_liu} leverages fully-connected and convolutional neural network (CNN) layers to enhance precoding performance.  A Bi-directional Long Short-Term Memory (Bi-LSTM) model is used in~\cite{implicit_2022chen} to exploit subband correlations, while~\cite{implicit_2024gao} enhances CNN-based architectures through cross-polarized antenna-aware data organization and instance batch normalization. However, these works primarily optimize network structures without fully utilizing valuable side information available at the BS, thereby limiting further improvement in compression efficiency.

To address this limitation, a recent state-of-the-art (SOTA) method~\cite{mixeddata2024_2} adapts bi-directional correlation—originally used for full CSI feedback~\cite{bifull_liu2019}—to implicit CSI scenarios. Since the uplink and downlink share similar propagation characteristics, such as angles of arrival/departure and delay spreads~\cite{bidirection_zhong2020}, FDD channels inherently exhibit a notable bi-directional correlation in the angular-delay domain \cite{bifull_liu2019}. 
Leveraging this, Ubi-ImCsiNet~\cite{mixeddata2024_2} employs uplink eigenvector magnitudes as side information during decoding, reducing the required feedback rate.
However, key challenges hinder the practical deployment of
bi-directional correlation-assisted ultra-low-rate implicit CSI
feedback. First, unlike the strong uplink-downlink correlation in full CSI feedback~\cite{bifull_liu2019}, the correlation in implicit CSI feedback can be degraded due to the non-uniqueness of eigenvectors~\cite{Horn_Johnson_1985}, which weakens CSI compression efficiency; second, DL-based models exhibit limited generalization capabilities, suffering from degraded performance when applied to unseen radio-frequency (RF) environments.

Motivated by these challenges, we propose a novel DL-based implicit CSI feedback framework, named \emph{Dual-ImRUNet}, aiming to achieve ultra-low-rate CSI feedback with strong robustness across varying RF environments. Specifically, Dual-ImRUNet integrates a novel bi-directional correlation enhancement preprocessing module with an efficient transformer-based implicit CSI feedback network, significantly reducing feedback overhead to single-digit bit counts. Besides, by proposing an innovative plug-in uplink-downlink input format alignment module, Dual-ImRUNet maintains robustness to environmental variations, thus enhancing practical viability.

The main contributions are summarized as follows:
\begin{itemize}
    \item We propose a novel bi-directional correlation enhancement preprocessing module to reduce redundancy in ultra-low-rate implicit CSI compression. By projecting highly correlated uplink and downlink channel matrices into their respective subband eigenspaces, this module effectively mitigates bi-directional correlation degradation caused by eigenvector non-uniqueness.

    \item We develop an efficient transformer-based implicit CSI feedback network that leverages bi-directional correlation and angular-delay domain sparsity. Simulation results demonstrate that our proposed network achieves an 85\% reduction in feedback overhead, compressing CSI to only 6 bits, with merely 32\% of model complexity and 1\% of model size compared to the current SOTA, Ubi-ImCsiNet.

    \item To overcome performance degradation in unseen environments, we introduce an innovative input format alignment module to ensure consistent data distribution without incurring additional transmission costs. By establishing paired uplink-downlink reference benchmarks, Dual-ImRUNet effectively aligns input formats at both encoder and decoder independently. Simulation results demonstrate that Dual-ImRUNet trained in a single environment achieves robust performance across 30 unseen environments.
\end{itemize}


\section{System Model and Problem Formulation}
\subsection{System Model}
We consider a massive MIMO system where the BS has $N_\text{Tx}$ transmit antennas and the UE has $N_\text{Rx}$ receive antennas. The system employs orthogonal frequency division multiplexing (OFDM) over $N_\text{s}$ subbands, each with $N_\text{gran}$ subcarriers. The downlink frequency-domain channel matrix is $\mathbf{H}^{*}_\text{DL} \in \mathbb{C}^{N_\text{Rx} \times N_\text{Tx} \times N_\text{s}}$, represented as
$\mathbf{H}^{*}_\text{DL} = [\mathbf{H}_{1,\text{DL}},\, \mathbf{H}_{2,\text{DL}},\, \ldots,\, \mathbf{H}_{N_\text{s},\text{DL}}],$
where $\mathbf{H}_{s,\text{DL}} \in \mathbb{C}^{N_\text{Rx} \times N_\text{Tx}}$ denotes the downlink channel matrix for the $s$-th subband. Similarly, the uplink channel matrix is $\mathbf{H}^{*}_\text{UL} \in \mathbb{C}^{N_\text{Rx} \times N_\text{Tx} \times N_\text{s}}$.

For implicit CSI feedback in 5G NR, only the eigenvectors of the channel matrices are required for precoding~\cite{3gpp2022}. Assuming single-stream downlink transmission—a common assumption in implicit CSI feedback studies~\cite{eigennet_liu, implicit_2022chen, mixeddata2024_2, implicit_2024gao}—we compute the normalized eigenvector $\mathbf{w}_{s,\text{DL}} \in \mathbb{C}^{N_\text{Tx} \times 1}$ corresponding to the largest eigenvalue $\lambda_{s,\text{DL}}$ of $\mathbf{H}_{s,\text{DL}}^\text{H} \mathbf{H}_{s,\text{DL}}$:
\begin{equation}
\mathbf{H}_{s,\text{DL}}^\text{H} \mathbf{H}_{s,\text{DL}}\, \mathbf{w}_{s,\text{DL}} = \lambda_{s,\text{DL}}\, \mathbf{w}_{s,\text{DL}},
\end{equation}
with $\| \mathbf{w}_{s,\text{DL}} \|^2 = 1$. Collecting all $N_\text{s}$ eigenvectors forms the downlink eigenvector matrix $\mathbf{W}_\text{DL} = [\mathbf{w}_{1,\text{DL}},\, \ldots,\, \mathbf{w}_{N_\text{s},\text{DL}}]^\text{T} \in \mathbb{C}^{N_\text{s} \times N_\text{Tx}}$. The corresponding uplink eigenvector matrix is denoted as $\mathbf{W}_\text{UL} \in \mathbb{C}^{N_\text{s} \times N_\text{Tx}}$.

\subsection{Problem Formulation}
\vspace*{-1mm}
Our optimization objective follows a general form in implicit CSI feedback~\cite{eigennet_liu, implicit_2024gao}:
\begin{equation}
\max_{\mathfrak{J}}\, \overline{\rho^2}(\mathbf{W}_\text{DL},\, \hat{\mathbf{W}}_\text{DL}) = \max_{\mathfrak{J}}\, \frac{1}{N_\text{s}} \sum_{s=1}^{N_\text{s}} \frac{ \left| \mathbf{w}_{s,\text{DL}}^\mathrm{H} \hat{\mathbf{w}}_{s,\text{DL}} \right|^2 }{ \| \mathbf{w}_{s,\text{DL}} \|^2\, \| \hat{\mathbf{w}}_{s,\text{DL}} \|^2 },
\end{equation}
where $\rho^2(\cdot,\cdot)$ is the squared generalized cosine similarity (SGCS) metric used in 3GPP TR 38.843~\cite{3gpp2023csi}, and $\mathfrak{J}$ denotes the model parameters. 


Let $f_\text{en}(\cdot;\, \Phi)$ and $f_\text{de}(\cdot;\, \Psi)$ denote the encoder and decoder networks, parameterized by $\Phi$ and $\Psi$, respectively. In an uplink-assisted implicit CSI feedback framework~\cite{mixeddata2024_2}, the operations are formulated as
\begin{align}
\mathbf{c} &= f_\text{en}(\mathbf{W}_\text{DL};\, \Phi), \\
\hat{\mathbf{W}}_\text{DL} &= f_\text{de}(\mathbf{c},\, |\mathbf{W}_\text{UL}|;\, \Psi),
\end{align}
where $|\cdot|$ represents element-wise magnitude and $\mathbf{c}$ denotes the compressed codeword.

To enable ultra-low-rate feedback while ensuring strong generalization, we introduce two preprocessing modules: a bi-directional correlation enhancement (BCE) function $f_\text{BCE}(\cdot)$ and an input format alignment (IFA) function $f_\text{IFA}(\cdot)$. Specifically, $f_\text{BCE}(\cdot)$ optimizes eigenvectors within their eigenspaces to strengthen the correlation between uplink and downlink matrices, while $f_\text{IFA}(\cdot)$ aligns the input samples to a predefined benchmark through operations such as circular shifting.

At the UE side, the encoding process is:
\begin{eqnarray}
\mathbf{W}_\text{BCE,DL} &=& f_\text{BCE}(\mathbf{W}_\text{DL}), \\
\mathbf{W}_\text{IFA,DL},\, \mathbf{b}_\text{DL} &=& f_\text{IFA}(\mathbf{W}_\text{BCE,DL},\, \mathbf{W}_\text{Ben,DL}), \\
\mathbf{c} &=& f_\text{en}(\mathbf{W}_\text{IFA,DL};\, \Phi),
\end{eqnarray}
where $\mathbf{W}_\text{BCE,DL}$ is the downlink eigenvector matrix after correlation enhancement, $\mathbf{W}_\text{IFA,DL}$ is the eigenvector matrix aligned with the benchmark $\mathbf{W}_\text{Ben,DL}$, and  $\mathbf{b}_\text{DL}$ contains auxiliary control information for original data format restoration.

 
At the BS side, the recovery process is:
\begin{eqnarray}
\mathbf{W}_\text{BCE,UL} &=& f_\text{BCE}(\mathbf{W}_\text{UL}), \\
\mathbf{W}_\text{IFA,UL},\, \mathbf{b}_\text{UL} &=& f_\text{IFA}(\mathbf{W}_\text{BCE,UL},\, \mathbf{W}_\text{Ben,UL}), \\
\hat{\mathbf{W}}_\text{IFA,DL} &=& f_\text{de}(\mathbf{c},\, |\mathbf{W}_\text{IFA,UL}|;\, \Psi), \\
\hat{\mathbf{W}}_\text{BCE,DL} &=& f_\text{IFA}^{-1}(\hat{\mathbf{W}}_\text{IFA,DL},\, \mathbf{b}_\text{UL}),
\end{eqnarray}
where $\hat{\mathbf{W}}_\text{IFA,DL}$ is the recovered downlink eigenvector matrix in the aligned format, and $\hat{\mathbf{W}}_\text{BCE,DL}$ is the recovered downlink eigenvector matrix with enhanced bi-directional correlation. The subscript ``UL'' denotes uplink matrices and control information, similar to the notation in Eqs.~(5)--(7).

Notably, $\mathbf{W}_\text{BCE,DL}$ and $\mathbf{W}_\text{DL}$ correspond to the same eigenvalues, thus ensuring the consistency of precoding performance after enhancement and recovery. $\mathbf{b}_\text{UL}$ is used for downlink format restoration by leveraging the bi-directional correlation\cite{bifull_liu2019,bifull_2024xu}, thereby avoiding the additional transmission cost for $\mathbf{b}_\text{DL}$.

\begin{figure}
      \centering
      \includegraphics[scale=0.25]{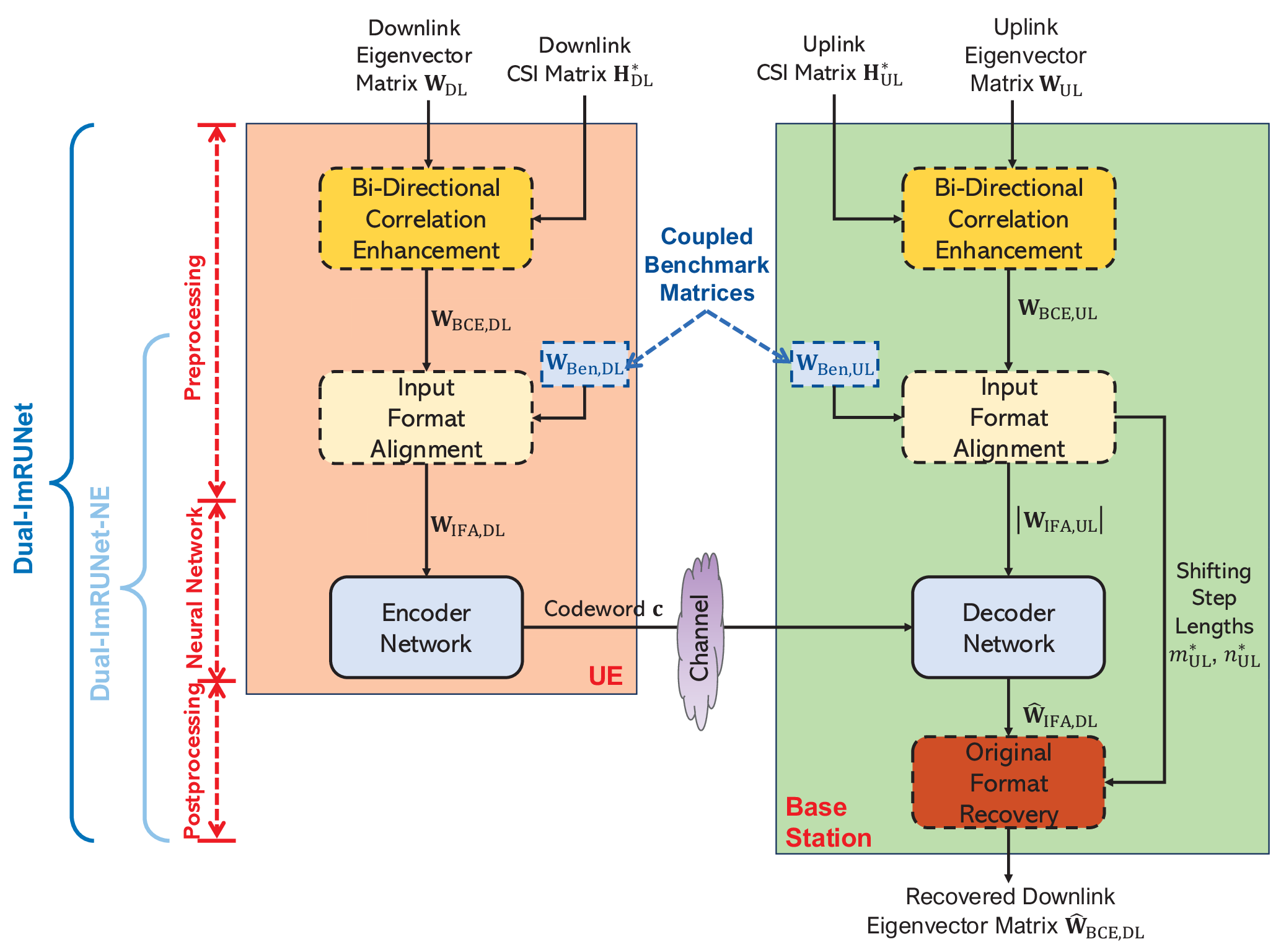}\vspace*{-2mm}
      \caption{Architecture of proposed uplink-assisted ultra-low-rate implicit CSI feedback framework.}
      \label{figure_architecture}
      \vspace*{-3mm}
  \end{figure}

\section{Proposed Implicit CSI Feedback Framework}

In this section, we present \emph{Dual-ImRUNet}, a novel uplink-assisted implicit CSI feedback framework designed to achieve ultra-low feedback rates while ensuring robust performance across diverse environments.   The framework consists of three
main components: a BCE
module, an IFA module, and a neural network for extreme compression and uplink-assisted recovery.

\subsection{Bi-Directional Correlation Enhancement}
\vspace*{-1mm}
Previous studies ~\cite{bifull_liu2019, bifull_2024xu} have leveraged uplink CSI at the BS to reduce redundancy and enhance recovery accuracy in full CSI feedback systems. However, in implicit CSI feedback, the correlation between uplink and downlink eigenvectors can be degraded due to the inherent non-uniqueness of eigenvector solutions, thereby limiting compression efficiency.

Specifically, for a channel matrix $\mathbf{H}_{s} \in \mathbb{C}^{N_\text{Rx} \times N_\text{Tx}}$ at subband $s$, the normalized eigenvector $\mathbf{w}_{s}$ corresponding to eigenvalue $\lambda_{s}$ is not unique \cite{Horn_Johnson_1985}. Any vector of the form $\mathbf{w}_{s} e^{j\phi}$ for $\phi \in \mathbb{R}$ also satisfies the eigen-equation. In cases where the eigenvalue $\lambda_{s}$ has a multiplicity greater than one, the corresponding eigenspace $\mathcal{E}_s$ may be of higher dimension, further exacerbating the bi-directional correlation degradation.

To address this, we propose a BCE function $f_\text{BCE}(\cdot)$, which optimizes the uplink and downlink eigenvector matrices $\mathbf{W}_\text{BCE,UL}$ and $\mathbf{W}_\text{BCE,DL}$ within their respective eigenspaces to maximize their mutual correlation.

A practical constraint arises from the absence of opposite link CSI at the UE and BS during feedback. Therefore, instead of jointly optimizing both sides, we adopt a suboptimal yet bandwidth-efficient strategy: using the original channel matrices $\mathbf{H}^{*}_\text{UL}$ and $\mathbf{H}^{*}_\text{DL}$, which are inherently correlated due to the shared propagation environment~\cite{bifull_liu2019,bidirection_zhong2020}, as references for enhancing eigenvector correlation.  

The bi-directional correlation enhancement thus can be formulated as two independent problems:
\begin{equation}
\mathbf{\tilde{w}}_{s,\text{DL}} = \underset{\mathbf{w}_{s,\text{DL}} \in \mathcal{E}_{s,\text{DL}}}{\operatorname{arg\,min}} \left\| \mathbf{w}_{s,\text{DL}} - \mathbf{h}_{s,\text{DL}} \right\|, \quad s = 1, \ldots, N_\text{s},
\label{downlink_enh}
\end{equation}
on the UE side, and
\begin{equation}
\mathbf{\tilde{w}}_{s,\text{UL}} = \underset{\mathbf{w}_{s,\text{UL}} \in \mathcal{E}_{s,\text{UL}}}{\operatorname{arg\,min}} \left\| \mathbf{w}_{s,\text{UL}} - \mathbf{h}_{s,\text{UL}} \right\|, \quad s = 1, \ldots, N_\text{s},
\end{equation}
on the BS side,
where $\|\cdot\|$ denotes the $\ell_2$-norm, $\mathcal{E}_{s,\text{DL/UL}}$ is the eigenspace corresponding to eigenvalue $\lambda_{s,\text{DL/UL}}$, and $\mathbf{h}_{s,\text{DL/UL}}$ is a reference vector derived from the channel matrix $\mathbf{H}_{s,\text{DL/UL}}$.

To illustrate the way to obtain the correlation-enhanced eigenvector matrix, we take Eq.~(\ref{downlink_enh}) as an example. Since $\mathbf{H}_{s,\text{DL}}^\text{T} \in \mathbb{C}^{N_\text{Tx} \times N_\text{Rx}}$ is larger than $\mathbf{w}_{s,\text{DL}} \in \mathbb{C}^{N_\text{Tx} \times 1}$, we choose $\mathbf{h}_{s,\text{DL}} = (\mathbf{H}_{s,\text{DL}})_{:, 1}^\text{T} \in \mathbb{C}^{N_\text{Tx} \times 1}$, representing the channel between the first receive antenna and all transmit antennas in subband $s$. The channel vector from any other receive antenna can also serve as $\mathbf{h}_{s,\text{DL}}$.

The solution involves projecting the reference vector onto the corresponding eigenspace. For the eigenvalue with a multiplicity of one, the adjusted eigenvector $\mathbf{\tilde{w}}_{s,\text{DL}}$ is obtained by:
\begin{equation}
\mathbf{\tilde{w}}_{s,\text{DL}} = \frac{\mathbf{w}_{s,\text{DL}}\mathbf{w}_{s,\text{DL}}^\text{H}\mathbf{h}_{s,\text{DL}}}{\|\mathbf{w}_{s,\text{DL}}\mathbf{w}_{s,\text{DL}}^\text{H}\mathbf{h}_{s,\text{DL}}\|}.
\end{equation}
 For higher multiplicities, $\mathbf{w}_{s,\text{DL}} \mathbf{w}_{s,\text{DL}}^\text{H}$ is replaced by the projection matrix formed by the orthonormal basis of $\mathcal{E}_{s,\text{DL}}$ \cite{Horn_Johnson_1985}.

 Finally, concatenating the eigenvectors $\mathbf{\tilde{w}}_{s,\text{DL}}$ across all $N_\text{s}$ subbands yields the optimized downlink eigenvector matrix $\mathbf{W}_{\text{BCE,DL}} = [\mathbf{\tilde{w}}_{1,\text{DL}}, \ldots, \mathbf{\tilde{w}}_{N_\text{s},\text{DL}}]^\text{T}$. Similarly, we can obtain the correlation-enhanced uplink eigenvector matrix $\mathbf{W}_{\text{BCE,UL}}$.

\begin{figure}[thpb]
\vspace*{-3mm}
    \centering
    \includegraphics[scale=0.36]{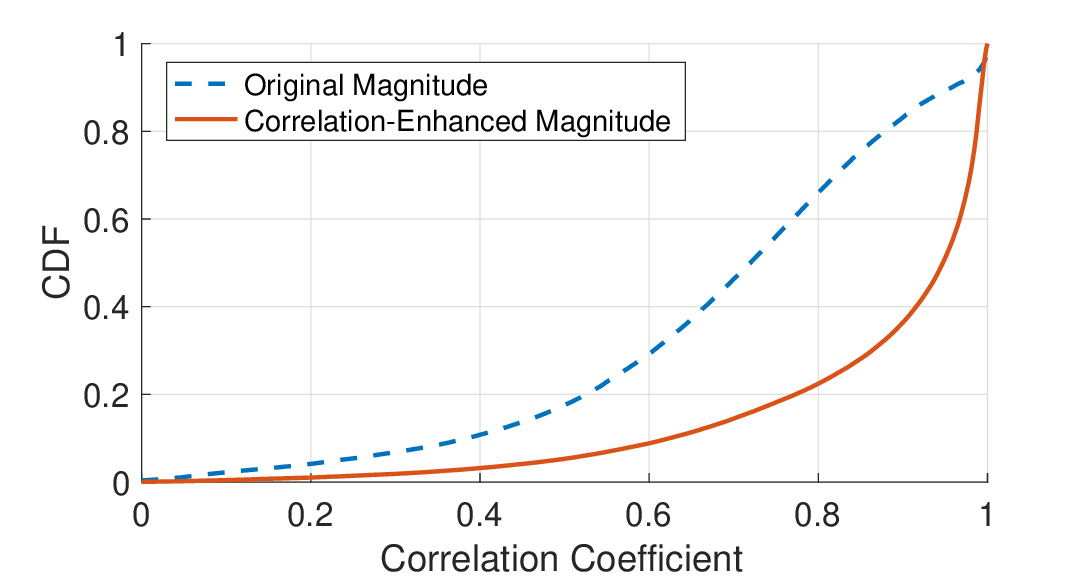}\vspace*{-1mm}
    \caption{Cumulative distribution of Pearson correlation coefficients between uplink and downlink eigenvector magnitudes in the studied massive MIMO system.}
    \label{figure_visual_cdf}
    \vspace*{-3mm}
\end{figure}

Fig.~\ref{figure_visual_cdf} shows the cumulative distribution function (CDF) of the Pearson correlation coefficients among the magnitudes of uplink and downlink eigenvector matrices. We use a ray-tracing-based wireless channel simulator~\cite{WirelessAI2022} with a downlink frequency of $2.6$ GHz and a central frequency gap of $120$ MHz; detailed settings are introduced in Section~IV.A. As shown in Fig.~\ref{figure_visual_cdf}, our proposed method significantly increases the correlation coefficients compared to conventional eigenvector decomposition, as evidenced by the CDF shifting towards higher values. The stronger correlation between the magnitudes of the uplink and downlink eigenvector matrices allows leveraging the enhanced uplink magnitude to more effectively reconstruct downlink CSI in ultra-low-rate implicit feedback.

\subsection{Input Format Alignment}

The performance of DL-based CSI feedback schemes often deteriorates in unseen environments due to input distribution mismatch. To mitigate this issue, we propose an input format alignment function $f_\text{IFA}(\cdot)$,  which aligns the input samples with a predefined benchmark to ensure consistency and compatibility with the pretrained CSI feedback model. By integrating the bi-directional correlation into the original format recovery, our design can recover the original format before alignment at the BS without extra transmission cost.

First, we implement input format alignment by transforming the eigenvector matrix into an approximate angular-delay domain
using the 2D Discrete Fourier Transform (DFT), resulting in sparsity~\cite{cui2022}. The eigenvector matrix $\mathbf{W}_\text{BCE,DL}$ becomes $\mathbf{W}_\text{Spar,DL}$:
\begin{equation}
	\mathbf{W}_\text{Spar,DL} = \mathcal{F}(\mathbf{W}_\text{BCE,DL}) = \mathbf{F}_\text{d}^\text{H} \mathbf{W}_\text{BCE,DL} \mathbf{F}_\text{a},
\end{equation}
where $\mathbf{F}_\text{d} \in \mathbb{C}^{N_\text{s} \times N_\text{s}}$ and $\mathbf{F}_\text{a} \in \mathbb{C}^{N_\text{Tx} \times N_\text{Tx}}$ are unitary DFT matrices, and $\mathcal{F}(\cdot)$ denotes the 2D DFT. The BS similarly computes the uplink sparse eigenvector matrix $\mathbf{W}_\text{Spar,UL}$.

\begin{figure}
    \centering
    \includegraphics[scale=0.28]{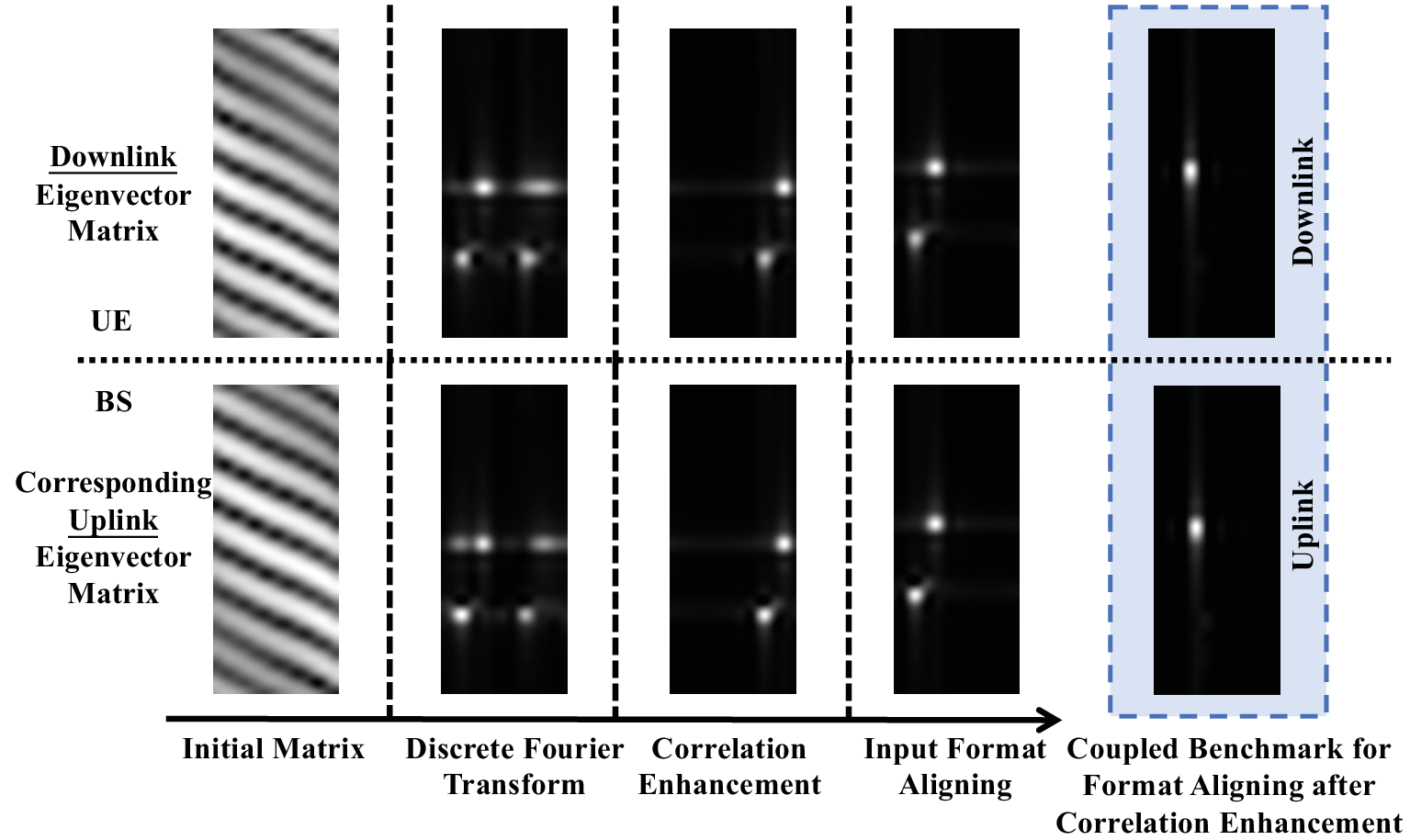}
    \caption{Visualization of eigenvector matrix magnitudes in the uplink and downlink under the influence of proposed modules.}
    \label{figure_visual}
    \vspace*{-3mm}
\end{figure}

Next, we construct coupled benchmark matrices $\{\mathbf{W}_\text{Ben,DL}, \mathbf{W}_\text{Ben,UL}\}$ deployed at the UE and BS, respectively, for alignment. We choose as benchmark matrices a pair of downlink and uplink matrices corresponding to a physical channel with only a Line-of-Sight (LoS) path. A visualization of the coupled benchmark matrices is shown on the right side of Fig.~\ref{figure_visual}. These benchmarks emphasize the primary component, which can be the LoS path in LoS scenarios or the strongest path in Non-Line-of-Sight (NLoS) scenarios. While more complex benchmark constructions may further improve recovery accuracy, they may also incur additional computational costs; we leave this exploration to future work.

To align $\mathbf{W}_\text{Spar,DL}$ with the benchmark $\mathbf{W}_\text{Ben,DL}$, we employ circular shifts using the function $f_\text{cs}(\cdot, m, n)$, where $m$ and $n$ are the shift steps along the rows and columns, respectively. 
The shifted eigenvector matrix is given by $\mathbf{W}_\text{cs,DL} = f_\text{cs}(\mathbf{W}_\text{Spar,DL}, m, n)$, computed as:
\begin{equation}
\begin{split}
 \mathbf{W}_{i,j,\text{cs,DL}} = \mathbf{W}_{(i + m) \bmod N_\text{Tx},\ (j + n) \bmod N_\text{s},\text{Spar,DL}}, \\ \forall\ i\in\{1,2,\dots,N_\text{Tx}\},\ j\in\{1,2,\dots,N_\text{s}\}.   
\end{split}
\end{equation}

The shift parameters $\mathbf{b}_\text{DL} = \{m^*_\text{DL}, n^*_\text{DL}\}$ are determined by maximizing the correlation between the row and column sums of $\mathbf{W}_\text{Spar,DL}$ and those of $\mathbf{W}_\text{Ben,DL}$. To simplify computation, \( m^*_\text{DL} \) and \( n^*_\text{DL} \) are solved independently.

Define the row sum vector of the magnitudes of $\mathbf{W}_\text{Spar,DL}$ as
$
\mathbf{r}_\text{DL}[i] = \sum_{j=1}^{N_\text{s}} \left| \mathbf{W}_{i,j,\text{Spar,DL}} \right|, \quad i = 1,2,\dots,N_\text{Tx}.
$
Similarly, the benchmark's row sum vector is defined as
$
\mathbf{r}_\text{Ben,DL}[i] = \sum_{j=1}^{N_\text{s}} \left| \mathbf{W}_{i,j,\text{Ben,DL}} \right|, \quad i = 1,2,\dots,N_\text{Tx}.
$

Next, we determine the optimal cyclic shift \( m^*_\text{DL} \) that maximizes the correlation between \( \mathbf{r}_\text{DL} \) and \( \mathbf{r}_\text{Ben,DL} \):
\begin{equation}
    C_\text{DL}(m) = \mathbf{r}_{\text{DL}}^{(m)} \cdot \mathbf{r}_\text{Ben,DL},
\end{equation}
where \( \mathbf{r}_{\text{DL}}^{(m)} \) is \( \mathbf{r}_\text{DL} \) after a cyclic shift by \( m \) steps. The optimal shift \( m^*_\text{DL} \) is given by:
\begin{equation}
    m^*_\text{DL} = \arg\max_m C_\text{DL}(m).
\end{equation}
Similarly, we obtain the optimal column shift \( n^*_\text{DL} \) by applying the same procedure to the column sum vectors.

The aligned downlink eigenvector matrix is then given by:
\begin{equation}
    \mathbf{W}_\text{IFA,DL} = f_\text{cs}(\mathbf{W}_\text{Spar,DL}, m^*_\text{DL}, n^*_\text{DL}),
\end{equation}
where $\{m^*_\text{DL}, n^*_\text{DL}\}$ are included in control information $\mathbf{b}_\text{DL}$.

At the BS, we similarly obtain \( \mathbf{b}_\text{UL} = \{m^*_\text{UL}, n^*_\text{UL}\} \) and \( \mathbf{W}_\text{IFA,UL} \) to align the uplink magnitude matrices.

Finally, to reconstruct the eigenvector matrix from the neural network's output $\hat{\mathbf{W}}_\text{IFA,DL}$, we apply the inverse of the format standardization process:
\begin{equation}
    \hat{\mathbf{W}}_\text{BCE,DL} = f_\text{cs}(\hat{\mathbf{W}}_\text{IFA,DL}, -m^*_\text{UL}, -n^*_\text{UL}),
\end{equation}
where $\mathcal{F}^{-1}(\cdot)$ denotes the inverse sparse transform function, specifically the inverse 2D DFT, enabling recovery of the original eigenvector matrix format.

Fig.~\ref{figure_visual} illustrates the effects of our proposed modules on the downlink and uplink eigenvector matrices. In these visualizations, brightness intensity reflects the magnitude of matrix elements, indicating sparsity and element strength.
Initially, the matrices show coarse periodic variations. After applying the 2D DFT, both matrices exhibit four bright clusters with differing shapes, indicating degraded correlation. The subsequent bi-directional correlation enhancement significantly increases the similarity between the uplink and downlink matrices, as shown in the third column of the figure. Finally, the input format alignment introduces a consistent sparsity pattern in both matrices by aligning the cluster with the strongest power to the same position, which can enable a more consistent precoding performance across varying communication environments.

\subsection{Encoder and Decoder Networks}

In this subsection, we propose a novel transformer-based, uplink-assisted implicit CSI feedback network to achieve ultra-low-rate feedback and high-accuracy recovery. 



\begin{figure}[thpb]
      \centering
      \includegraphics[scale=0.3]{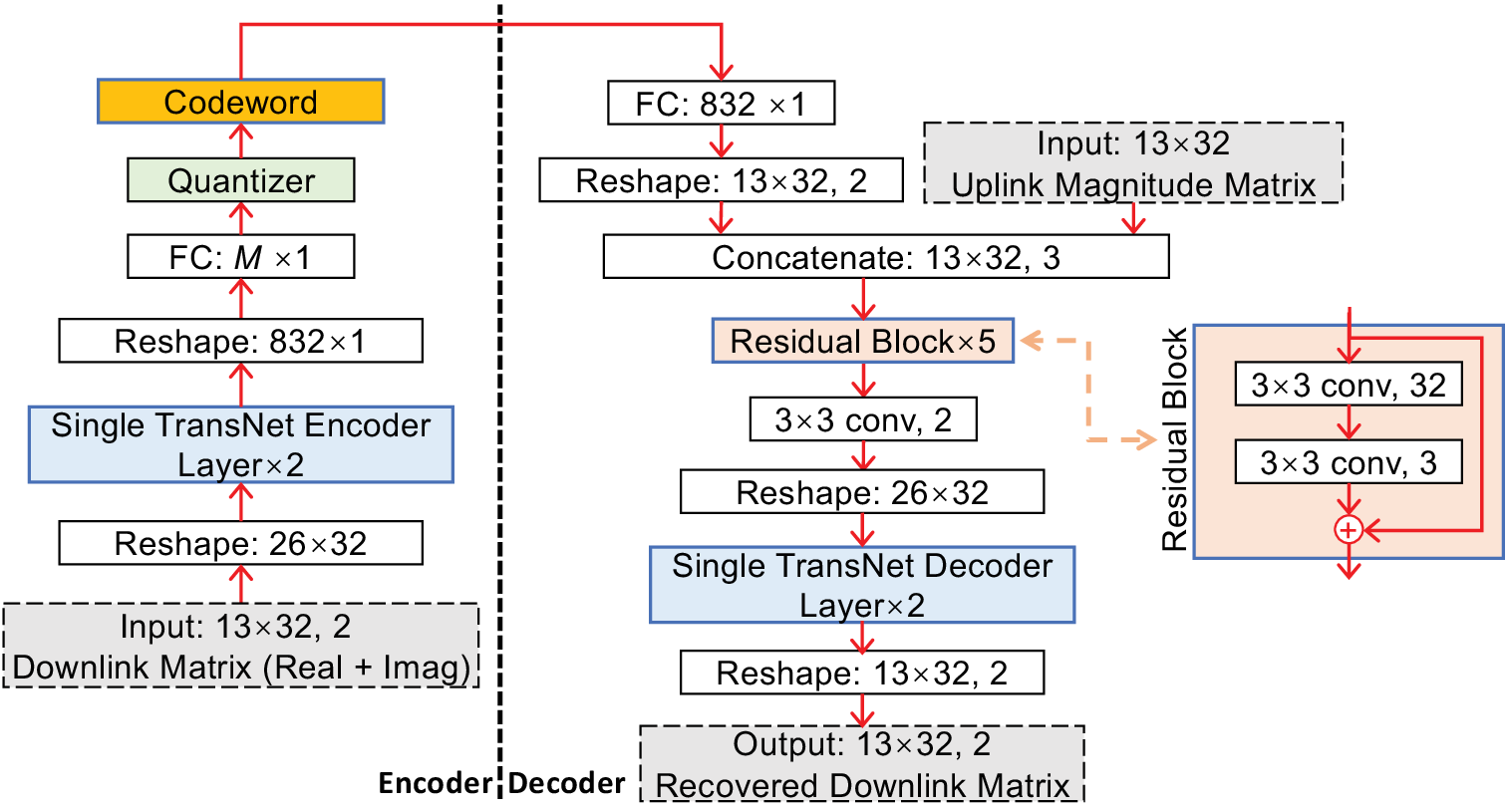}
      \caption{The backbone network structure of Dual-ImRUNet.}
      \label{figure_architecture_dual}
      \vspace*{-5mm}
\end{figure}

As shown in Fig.~\ref{figure_architecture_dual}, Dual-ImRUNet employs an autoencoder architecture. The encoder separates the complex eigenvector matrix into real and imaginary parts, which are then concatenated along the subband axis to form a feature map. Two transformer-based \emph{Single TransNet Encoder Layers} \footnote{Due to the space limitation, the detailed structure of \emph{Single TransNet Encoder Layers} can be found in ~\cite{cui2022}.} are applied for feature extraction, followed by a fully connected layer with \( M \) units for dimensionality reduction. Subsequently, a quantizer module~\cite{quan_liu} is used to quantize the dimension-compressed floating-point vector.

At the BS, the decoder reconstructs the downlink eigenvector matrix using the compressed codewords and the uplink magnitude matrix. The codewords are restored to their original length via a fully connected layer. A conjugation layer combines the reshaped downlink features with the uplink magnitude for decoding. To enhance reconstruction accuracy by leveraging uplink-downlink correlation, we employ five residual blocks~\cite{He_2016_CVPR}, each containing two \( 3 \times 3 \) convolutional layers with 32 and 2 channels, respectively. A \( 3 \times 3 \) convolutional layer then reduces feature maps from 3 to 2, followed by a reshape layer to concatenate the feature maps into one. Two transformer-based \emph{Single TransNet Decoder Layers}~\cite{cui2022} are used for fine-grained recovery. Finally, a reshape layer separates the real and imaginary parts of the recovered downlink eigenvector matrices.

Different from Ubi-ImCsiNet~\cite{mixeddata2024_2}, which uses uplink magnitudes to refine the output of bi-ImCsiNet~\cite{implicit_2022chen} that relies solely on the compressed codeword for recovery—potentially degrading performance if the initial recovery is inaccurate—our approach incorporates the uplink magnitudes after dimension recovery. This allows for full leverage of the bi-directional channel correlation, effectively compensating for missing downlink magnitude information and enabling ultra-low-rate feedback. Furthermore, whereas Ubi‐ImCsiNet’s use of Bi‐LSTM layers to capture inter‐subband dependencies inherently restricts the minimum codeword length to the number of subbands, Dual‐ImRUNet’s transformer‐based encoder compresses the feedback down to a single scalar while maintaining high feedback accuracy.

Our framework, combining BCE, IFA, and the neural network, is called \emph{Dual-ImRUNet}. The version without BCE is referred to as \emph{Dual-ImRUNet-NE}.

\section{Performance Evaluation}
\subsection{Simulation Setting}
We evaluate the performance using a ray-tracing-based wireless channel simulator~\cite{WirelessAI2022}, generating CSI data from 100 real-world city maps. Each map provides channel information between a BS and 10,000 UE locations. Simulation parameters are summarized in Table~I, with BS and UE equipped with uniform linear arrays\cite{mixeddata2024_2} and a center band gap of 120 MHz, consistent with the band gap at 2.6 GHz in~\cite{3gpp2024band}.

\begin{table}
\centering
\vspace*{3mm}
\caption{Basic Simulation Parameters}
\begin{tabular}{|l|c|}
\hline
\textbf{Parameter} & \textbf{Value} \\
\hline
System bandwidth & 10 MHz \\
\hline
Downlink center frequency & 2.60 GHz \\
\hline
Uplink center frequency & 2.48 GHz \\
\hline
Number of subbands \(N_\text{s}\) & 13 \\
\hline
Subband granularity \(N_\text{gran}\) & 48 \\
\hline
Transmitting antennas \(N_\text{Tx}\) & 32 \\
\hline
Receiving antennas \(N_\text{Rx}\) & 4 \\
\hline
\end{tabular}
\end{table}

Training, validation, and testing sets contain 112,000, 28,000, and 60,000 sample pairs, respectively. The testing set includes 2,000 samples from each of the last 30 unseen environments. To examine the impact of training diversity, the training and validation samples are sourced from the first 1 to 70 environments. If the initial dataset is insufficient, it is augmented through repetition to match the desired size; if too large, it is uniformly sampled. Models are trained with a batch size of 64, undergoing 1,000 training epochs without the quantizer and 100 epochs of fine-tuning with the quantizer. The quantization bits per element are set to $B = 6$.

We compare Dual-ImRUNet and Dual-ImRUNet-NE with two other CSI feedback schemes: Ubi-ImCsiNet~\cite{mixeddata2024_2} and TransNet~\cite{cui2022}. Ubi-ImCsiNet achieves excellent implicit CSI feedback performance by employing Bi-LSTM layers to capture subband correlations for codeword compression and supports uplink-assisted recovery. TransNet uses the same transformer structures in the encoder and decoder networks as Dual-ImRUNet but lacks uplink information for decoding. Although incorporating temporal correlation can provide additional gains~\cite{mixeddata2024_2}, it introduces new challenges in model efficiency and robustness, which we leave for future work.

Due to Ubi-ImCsiNet's minimum compressed code dimension of $N_s=13$, we trained it at compressed dimensions of 13, 26, and 39 with quantization bits per element from 1 to 6, selecting the best performance for each total bit count to show its performance in ultra-low-rate feedback. TransNet uses the same quantization settings as Dual-ImRUNet.

\subsection{Performance Comparison and Complexity Evaluation}

\begin{figure} 
\centering
\vspace*{-5mm}
\includegraphics[width=0.88\columnwidth]{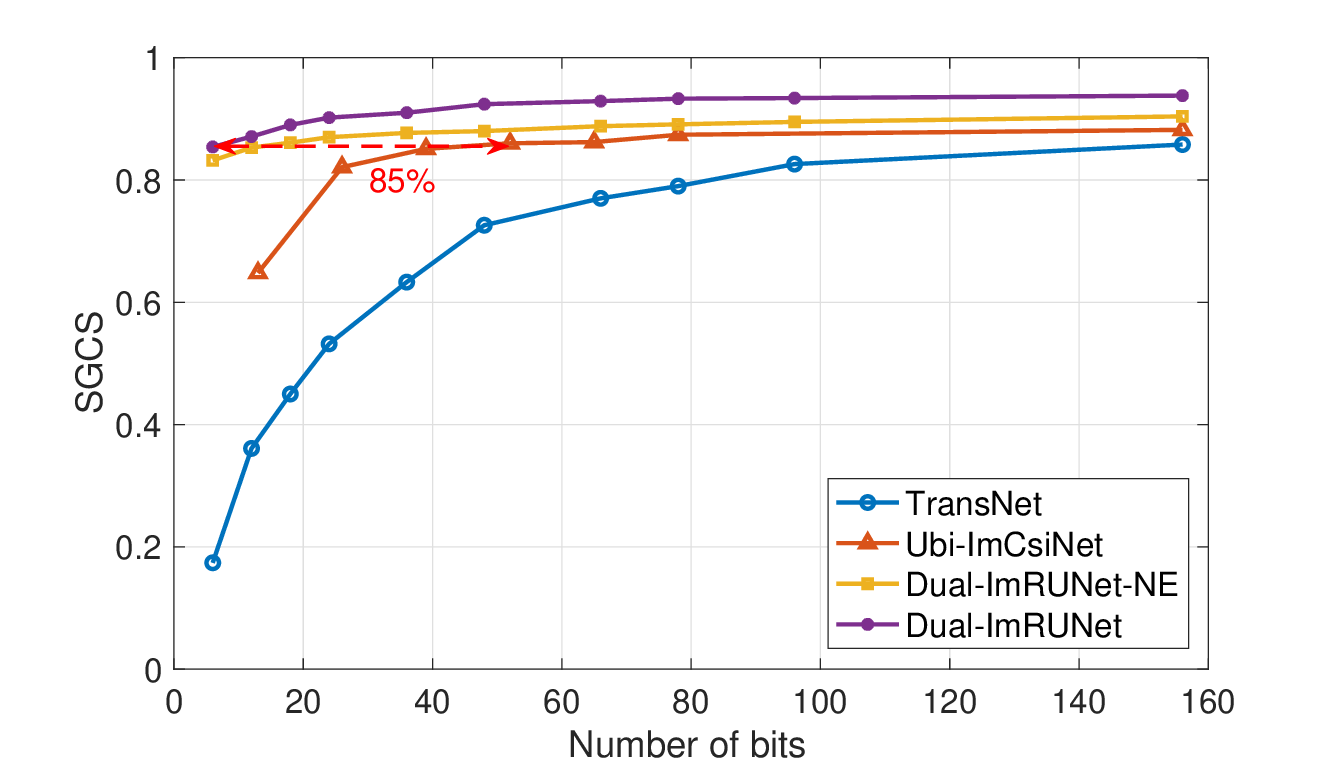}
\vspace*{-1mm}
\caption{CSI recovery comparison at different feedback bits when the training set is collected from 70 environments.} 
\vspace*{-5mm}
\label{figurepe2} 
\end{figure}

Fig.~\ref{figurepe2} illustrates the impact of feedback bits on CSI recovery in unseen environments when the training set is collected from 70 environments, i.e., with rich training diversity. As shown in Fig.~\ref{figurepe2}, Dual-ImRUNet and Dual-ImRUNet-NE demonstrate superior SGCS performance, especially when the feedback bits are below 20, confirming the effectiveness of our preprocessing modules and neural network structure optimization for the ultra-low-rate feedback. Moreover, Dual-ImRUNet outperforms Dual-ImRUNet-NE, validating the effectiveness of the proposed BCE module to enhance the bi-directional correlation for higher recovery accuracy. Besides, the performance gap between TransNet and Dual-ImRUNet is minimal at 156 feedback bits, indicating that most of the information is already in the codeword at high feedback rates.

Notably, compared with the SOTA Ubi-ImCsiNet,
Dual-ImRUNet can reduce the feedback overhead by 85\% to only 6 bits while achieving the SGCS of 0.85, highlighting a promising way for ultra-low-rate implicit CSI feedback.

\begin{figure}[!t]
\centering
\includegraphics[width=0.836\columnwidth]{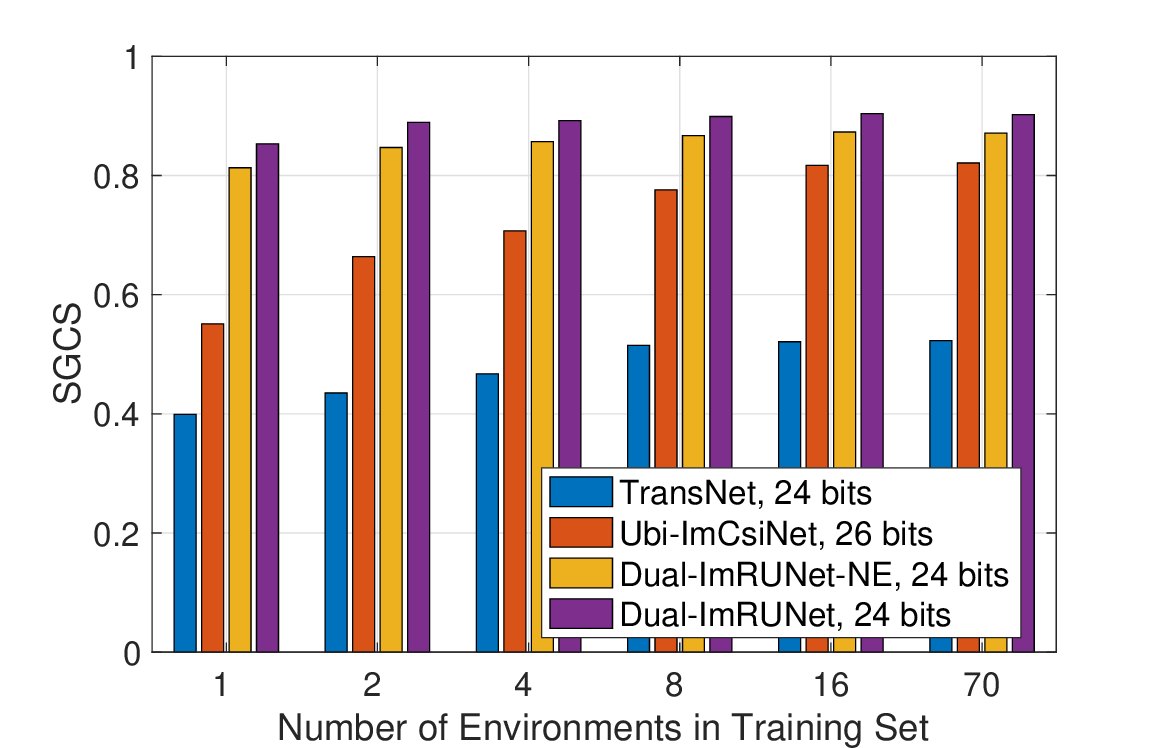}
\vspace*{-2mm}
\caption{CSI recovery comparison across a varying number of training environments, with feedback bits set around 24.}
\vspace*{-2mm}
\label{figure_envs}
\end{figure}

Fig.~\ref{figure_envs} illustrates performance at approximately 24 feedback bits across 30 unseen environments under varying training diversity. Dual-ImRUNet and Dual-ImRUNet-NE consistently outperform both baselines across all training diversity. As training diversity decreases, the performance gap between Dual-ImRUNet and Ubi-ImCsiNet increases. When trained on data from a single environment, Dual-ImRUNet achieves a 55\% SGCS improvement over Ubi-ImCsiNet and performs comparably to a model trained on data from 70 environments, highlighting the effectiveness of our IFA module to narrow the gap of input distributions among different environments.

\begin{table}[]
    \renewcommand{\arraystretch}{1.5}
    \centering
    \caption{Comparison of model size and computational complexity for different feedback bits. FLOPs: floating point operations, M: million, K: thousand, DIRUNet: Dual-ImRUNet.}
    \vspace*{-0.5mm}
    \label{tab:comp-complex}
    \begin{tabular}{|c|c|c|c|c|c|}
        \hline
        & \textbf{Bits} & \textbf{TransNet} & \textbf{Ubi-ImCsiNet} & \textbf{DIRUNet-NE} & \textbf{DIRUNet} \\
        \hline
        \multirow{3}{*}{\raisebox{-0.8\height}{\rotatebox{90}{\textbf{Parameters}}}} 
        & \textbf{$6$} & 111.11 K & N/A & 125.15 K & 127.58 K \\
        \cline{2-6}
        & \textbf{$24/26$} & 116.10 K & 15.87 M & 130.15 K & 132.58 K \\
        \cline{2-6}
        & \textbf{$156$} & 152.73 K & 16.31 M & 166.78 K & 169.21 K \\
        \hline
        \multirow{3}{*}{\raisebox{-0.9\height}{\rotatebox{90}{\textbf{FLOPs}}}} 
        & \textbf{$6$} & 5.58 M & N/A & 10.24 M & 10.26 M \\
        \cline{2-6}
        & \textbf{$24/26$} & 5.58 M & 31.65 M & 10.25 M & 10.27 M \\
        \cline{2-6}
        & \textbf{$156$} & 5.62 M & 32.53 M & 10.29 M & 10.31 M \\
        \hline
    \end{tabular}
    \vspace*{-5mm}
\end{table}

Table~\ref{tab:comp-complex} compares model size and computational complexity. Among the tested feedback bits, Dual-ImRUNet uses only 32\% of the FLOPs of Ubi-ImCsiNet and reduces parameters to 1\% while achieving better recovery performance. Additionally, Dual-ImRUNet introduces less than a 2\% increase in model size and less than a 0.2\% increase in FLOPs compared to Dual-ImRUNet-NE, as the bi-directional correlation enhancement adds negligible overhead relative to the neural networks.

\section{Conclusion}

This paper presents an uplink-assisted DL-based framework enabling robust, ultra-low-rate implicit CSI feedback in massive MIMO systems. By enhancing bi-directional correlation via eigenspace projection and employing an efficient transformer-based feedback network, our approach significantly reduces feedback overhead while maintaining high CSI recovery accuracy. Additionally, the bi-directional correlation-assisted input format alignment ensures consistent data representation across diverse environments and maintains robust performance. Simulation results show that Dual-ImRUNet can reduce the feedback overhead by 85\% to only 6 bits compared with the SOTA while preserving environmental robustness.


\ifCLASSOPTIONcaptionsoff
  \newpage
\fi

\bibliographystyle{IEEEtran}
\bibliography{ref}

\begin{thebibliography}{10}
\providecommand{\url}[1]{#1}
\csname url@samestyle\endcsname
\providecommand{\newblock}{\relax}
\providecommand{\bibinfo}[2]{#2}
\providecommand{\BIBentrySTDinterwordspacing}{\spaceskip=0pt\relax}
\providecommand{\BIBentryALTinterwordstretchfactor}{4}
\providecommand{\BIBentryALTinterwordspacing}{\spaceskip=\fontdimen2\font plus
\BIBentryALTinterwordstretchfactor\fontdimen3\font minus \fontdimen4\font\relax}
\providecommand{\BIBforeignlanguage}[2]{{%
\expandafter\ifx\csname l@#1\endcsname\relax
\typeout{** WARNING: IEEEtran.bst: No hyphenation pattern has been}%
\typeout{** loaded for the language `#1'. Using the pattern for}%
\typeout{** the default language instead.}%
\else
\language=\csname l@#1\endcsname
\fi
#2}}
\providecommand{\BIBdecl}{\relax}
\BIBdecl

\bibitem{3gpp2022}
{3GPP}, ``Physical layer procedures for data ({Release} 17),'' Technical Specification TS 38.214 V17.0.0, Jan. 2022.

\bibitem{3gpp2023csi}
------, ``Study on artificial intelligence/machine learning for {NR} air interface ({Release} 18),'' Technical Report TR 38.843 V2.0.1, Feb. 2024.

\bibitem{cui2022}
Y.~Cui, A.~Guo, and C.~Song, ``Transnet: Full attention network for {CSI} feedback in {FDD} massive {MIMO} system,'' \emph{IEEE Wireless Commun. Lett.}, vol.~11, no.~5, pp. 903--907, May 2022.

\bibitem{full_wang2023}
B.~Wang, Y.~Teng, V.~Lau, and Z.~Han, ``{CCA-Net}: A lightweight network using criss-cross attention for {CSI} feedback,'' \emph{IEEE Commun. Lett.}, vol.~27, no.~7, pp. 1879--1883, Jul. 2023.

\bibitem{bifull_liu2019}
Z.~Liu, L.~Zhang, and Z.~Ding, ``Exploiting bi-directional channel reciprocity in deep learning for low rate massive {MIMO} {CSI} feedback,'' \emph{IEEE Wireless Commun. Lett.}, vol.~8, no.~3, pp. 889--892, Jun. 2019.

\bibitem{bifull_2024xu}
W.~Xu, J.~Wu, S.~Jin, X.~You, and Z.~Lu, ``Disentangled representation learning empowered {CSI} feedback using implicit channel reciprocity in {FDD} massive {MIMO},'' \emph{IEEE Trans. on Wireless Commun.}, vol.~23, no.~10, pp. 15\,169--15\,184, 2024.

\bibitem{eigennet_liu}
W.~Liu, W.~Tian, H.~Xiao, S.~Jin, X.~Liu, and J.~Shen, ``{EVCsiNet}: Eigenvector-based {CSI} feedback under {3GPP} link-level channels,'' \emph{IEEE Wireless Commun. Lett.}, vol.~10, no.~12, pp. 2688--2692, 2021.

\bibitem{implicit_2022chen}
M.~Chen, J.~Guo, C.-K. Wen, S.~Jin, G.~Y. Li, and A.~Yang, ``Deep learning-based implicit {CSI} feedback in massive {MIMO},'' \emph{IEEE Trans. on Commun.}, vol.~70, no.~2, pp. 935--950, Feb. 2022.

\bibitem{implicit_2024gao}
R.~Gao, X.~Li, and W.~Chen, ``{DARENet}: Data arrangement neural network for eigenvector-based {CSI} feedback,'' \emph{IEEE Wireless Commun. Lett.}, vol.~13, no.~8, pp. 2215--2219, Aug. 2024.

\bibitem{mixeddata2024_2}
C.~Jiang, J.~Guo, C.~Wen, and S.~Jin, ``Multi-domain correlation-aided implicit {CSI} feedback using deep learning,'' \emph{IEEE Trans. on Wireless Commun.}, vol.~23, no.~10, pp. 13\,344--13\,358, Oct. 2024.

\bibitem{bidirection_zhong2020}
Z.~Zhong, L.~Fan, and S.~Ge, ``{FDD} massive {MIMO} uplink and downlink channel reciprocity properties: Full or partial reciprocity?'' in \emph{Proc. IEEE Global Commun. Conf.}, 2020, pp. 1--5.

\bibitem{Horn_Johnson_1985}
R.~A. Horn and C.~R. Johnson, \emph{Matrix Analysis}.\hskip 1em plus 0.5em minus 0.4em\relax Cambridge University Press, 1985, pp. 33--64.

\bibitem{WirelessAI2022}
{Mobile AI Dataset}, ``Wireless {AI} research dataset,'' Online, Nov. 2023, available: https://www.mobileai-dataset.cn/html/default/yingwen/DateSet/1590994253188792322.html? index=1\&language=en.

\bibitem{quan_liu}
Z.~Liu, L.~Zhang, and Z.~Ding, ``An efficient deep learning framework for low rate massive {MIMO} {CSI} reporting,'' \emph{IEEE Trans. on Commun.}, vol.~68, no.~8, pp. 4761--4772, Aug. 2020.

\bibitem{He_2016_CVPR}
K.~He, X.~Zhang, S.~Ren, and J.~Sun, ``Deep residual learning for image recognition,'' in \emph{Proc. IEEE Conf. Comput. Vis. Pattern Recognit.}, 2016, pp. 770--778.

\bibitem{3gpp2024band}
{3GPP}, ``Evolved universal terrestrial radio access ({E-UTRA}); user equipment ({UE}) radio transmission and reception ({Release} 18),'' Technical Specification TS 36.101 V18.7.0, Sep. 2024.

\end{thebibliography}

\end{document}